\documentclass[11pt]{article}
\usepackage{blois,epsfig,units}

\begin{document}
\vspace*{4cm}
\title{BARYONIC ACOUSTIC OSCILLATIONS SIMULATIONS FOR THE LARGE SYNOPTIC SURVEY TELESCOPE (LSST)}

\author{ A. Gorecki, A.Barrau, S.Baumont} 

\address{Laboratoire de Physique des Particules et de Cosmologie (UJF/INPG/CNRS/IN2P3)\\
  53 avenure des Martyrs, 38026 Grenoble Cedex, FRANCE}

\author{A.Abate, A.Ansari and M.Moniez }
\address{Laboratoire de l'Accélérateur Linéaire, Université Paris-Sud 11\\91898 Orsay cedex}

\maketitle\abstracts{The baryonic acoustic oscillations are features in the spatial distribution of the galaxies which, if observed at different epochs, probe the nature of the dark energy. In order to be able to measure the parameters of the dark energy equation of state  to high precision, a huge sample of galaxies has to be used. The Large Synoptic Survey Telescope will survey the optical sky with 6 filters from $\unit[300]{nm}$ and $\unit[1100]{nm}$, such that a catalog of galaxies with photometric redshifts will be available for dark energy studies. In this article, we will give a rough estimate of the impact of the photometric redshift uncertainties on the computation of the dark energy parameter through the reconstruction of the BAO scale from a simulated photometric catalog.
 }

\section{Introduction}
Before the time of recombination, the competition between the photon pressure that tends to drive the baryons away from the dark matter potential wells, and the gravitational attraction between the baryons and the dark matter, induced acoustic waves in the photon-baryon plasma. We call these acoustic waves the Baryonic Acoustic Oscillations (BAO). When the photons and the baryons are decoupled the oscillations forze until the baryon-drag epoch. The CMB temperature anisotropy peaks are directly related to the sound horizon ($r_s(z_d) = \unit[100]{h^{-1}Mpc}$ comoving) at the decoupling epoch ($z_d = 1100$). Similarly, the correlation function of the galaxies, that is supposed to trace the matter density field, exhibits a bump at the BAO scale, which, by taking the Fourier transform,  translates into oscillations ($k_{BAO}$) in the matter power spectrum. As time evolves, the Universe expands according to the equations of state of its different constituents so the angular diameter distance as a function of redshift, which is related to the comoving BAO wavelength, also evolves. We propose a simulation that should allow an estimation of the uncertainties of the dark energy equation of state parameters as they should be measured with the telescope LSST \cite{LSST:2009pq}, by reconstructing $k_{BAO}$ at different redshifts and estimating its uncertainties.

\section{Simulation pipeline}
\subsection{Fiducial cosmology}
We start by setting the value of the cosmological parameters, roughly the same as the WMAP-5 values for a $\Lambda CDM$ cosmology:  the dark matter density $\Omega_{m} = 0.27$, the dark energy density $\Omega_{\Lambda} = 0.73$, the baryon density $\Omega_{b} = 0.04$, the Hubble parameter $h = 0.71$, the dark energy equation of state $w = -1$, the spectral index $n = 1$ and the normalisation of the density fluctuations $\sigma_{8} = 0.8$.

\subsection{Matter power spectrum, density field and galaxy number}
 The matter power spectrum for a redshift $z$, $P(k,z)$ is computed with a model linear theory for the fiducial parameter values, from the fitting formulae of Eisenstein et al \cite{Hu:1998tj} and is scaled to $z$ using the approximation for the growth function $G(z)$ from Carroll et al \cite{Carroll:1991mt}:
\begin{eqnarray*}
	G(a) = \frac{5\Omega_m}{2}\frac{H(a)}{H_{0}}\int_0^a \frac{da'}{(a'H(a')/H_0)^3},\\
	P(k,z) = G(z)^2P(k,0),
\end{eqnarray*}
where $P(k,0)$ is the matter power spectrum measured today. A cuboid of Fourier coefficients is generated from $P(k,z)$ then Fast-Fourier Transformed into a cuboid of over-densities, $\delta(\textbf{r}) =(\rho(\textbf{r}) - \rho)/\bar{\rho}$, where $\rho(\textbf{r})$ is the matter density, and $\bar{\rho}$ is the background mean matter density. The matter density field relates to the number of galaxies in a cell of center $\textbf{r}$ and volume $V_c$ with: $N_i(\textbf{r}) = (\delta(\textbf{r}) + 1) V_c \times n$ where $n$ is the mean number density of galaxies per unit volume. We determine this density $n$ by integrating the luminosity function (LF) $\phi(M)$ that gives the probability of the expected number density of galaxies per unit volume and unit of absolute magnitude $M$:
\begin{eqnarray*}
	n = \int \phi(M)dM.	
\end{eqnarray*}
The LF is a Schechter parametric function, and the values of the parameters are the ones measured by the GOODS survey from Dahlen et al \cite{Dahlen:2005}. It is, in fact, a function of the type of the galaxy: \textit{early},  \textit{late} and  \textit{starburst} and is measured in the $B$ filter of the GOODS survey.

\subsection{Observed galaxy catalog}
Each of the simulated galaxies is characterized by an absolute magnitude $M_B$ and a spectral energy distribution (SED), that is fixed by the LF. The SED is taken to be one of 50 SED that are interpolated among the four CWW templates from Coleman et al \cite{1980ApJS...43..393C} and two starburst SED from Kinney et al \cite{1996ApJ...467...38K}. The apparent magnitudes in the 6 LSST bands $(u,g,r,i,z,y)$ are then computed via:
\begin{eqnarray*}
	m_X = M_B + \mu(z) + K_{BX}(z),
\end{eqnarray*}
where $X$ is a LSST filter, $\mu$ is the distance modulus, $z$ the galaxy redshift, and $K_{BX}$ is the K-correction that adjusts the magnitude from the rest-frame $B$ band of GOODS to the observed-frame $X$ band. The K-corrections are computed following the formulae in Hogg et al \cite{Hogg:2002yh}. A Gaussian photometric uncertainty is added to $m_X$ and is equal to:
\begin{eqnarray*}
	\sigma(m_X) = \sqrt{(0.04 - \gamma)10^{0.4(m_X - m_{5,X})}+ \gamma 10^{0.8(m_X - m_{5,X})}},
\end{eqnarray*}
where $m_{5,X}$ is the $5\sigma$ detection magnitude in the $X$ band. This equation and the form of $m_{5,X}$ are given in the LSST Science Book\cite{LSST:2009pq}. A systematic uncertainty is added in quadrature, and is set to 0.005. If $m_X>m_{5,X}$ the galaxy is not observed in this band. 

\subsection{Photometric redshift}
For each galaxy, the photometric redshift is computed using our own template fitting method based on a chi square minimization, that is very similar to the public code LEPHARE or HYPERZ \cite{Bolzonella:2000js}. A prior based on the luminosity function that disfavors the likelihood of obtaining a high redshift for high observed flux is used, such that the function to be minimized according to the redshift and the SED type $T$ is: 
\begin{eqnarray*}
	\chi^2(z, T) = 
	\displaystyle\sum_{X = 1}^{6} \left(\frac{ F_{obs}^{X} - MF_{exp}^{X}(z,T)}{\sigma_{ F_{obs}^{X} }} \right)^2 - 2\log(P(M_B|T,z)),
\end{eqnarray*}
where $ F_{obs}$ is the observed flux, $ F_{exp}$ is the expected flux, $M$ a normalization factor, and $P(M_B|T,z)$ is the prior based on the luminosity function.

\subsection{Power spectra and dark energy parameter reconstruction}
 A fiducial cosmology is assumed and the comoving distance to each galaxy is computed, where the same values of the cosmological parameters as the ones used to generate the observed galaxy catalog are taken. We lay a grid over the simulated galaxy survey and count the number of galaxies in each grid cell $n_g$. The galaxy over-density distribution in cells is then:
\begin{eqnarray*}
	\delta_g(\textbf{r}) = \frac{n_g(\textbf{r}) - \bar{n}}{\bar{n}},
\end{eqnarray*}
where $\bar{n}$ is the mean number density of galaxies, calculated from the total number of galaxies in the simulation divided by the simulation volume. The gridded over-density is then Fourier transformed, and the power spectrum is computed as a function of redshift slice. The comoving BAO scale is computed using the \textit{wiggles-only} method. The power spectrum is divided by a smooth, wiggle-free power spectrum, and the resulting function is well approximated by an empirical slowly decaying sinusoidal function with parameters $k_{BAO}$, $A$ the amplitude, and a decay parameter $\alpha$:
\begin{eqnarray}\label{Eq:PS}
	P^s_{sm} = \frac{P(k)}{P_{ref}} = 1 + Ak \exp\left[ -\left( \frac{k}{\unit[0.1]{Mpc^{-1}}}\right)^{\alpha} \right] \sin\left( \frac{2\pi k}{k_{BAO}} \right),
\end{eqnarray}
where $\alpha$ is set to 1.4 (see Eisenstein et al \cite{Hu:1998tj}). A chi square minimization is done by varying the parameter $A$ and $k_{BAO}$: 
\begin{eqnarray*}
	\chi^2(k_{BAO},A) = \sum_{k} \frac{\left(P^0_{sm}(k) - P^s_{sm}(k,k_{BAO},A)\right)^2}{\sigma(P^0_{sm}(k) )^2},
\end{eqnarray*}
where $P^0_{sm}(k)$ is the measured power spectrum divided by the wiggle-free reference spectrum $P_{ref}$ and $P^s_{sm}(k,k_{BAO},A)$ is given by Eq~\ref{Eq:PS}. The uncertainty on $\hat{k}_{BAO}$ is given by the solving $\chi^2(\hat{k}_{BAO}\pm\sigma(\hat{k}_{BAO}),\hat{A}\pm \sigma(\hat{A})) - \chi^2_{min} = 2,3$ at $68\%$ confidence level, where the hat stands for the values of the parameters that minimize the chi square.
The angular BAO scale $\theta_s$ is related to the comoving BAO wavelength and the angular diameter distance $D_A(w,z)$ at a redshift slice $z$ :
\begin{eqnarray}
	\theta_s = \frac{2\pi}{\hat{k}_{BAO}D_A(w,z)},
\end{eqnarray}
The CMB allows for a calibration of the BAO scale and gives $\theta_s$, therefore we constrain the value of $D_A(w,z)$. The angular diameter distance depends on the cosmological parameters, such that:
\begin{eqnarray}
	D_A(w,z) = \frac{c}{H_0}\int_0^z\frac{dz'}{\sqrt{\Omega_m(1+z')^3+\Omega_{\Lambda}(1+z')^{3(1+w)}}}
\end{eqnarray}
The relative uncertainty on $k_{BAO}$ is equal to the relative uncertainty on $D_A(w,z)$ which gives the uncertainty on $w$ at redshift $z$ as shown in Fig~\ref{Fig:Daw} on the right-hand panel.

\section{Results}
\begin{figure}[]
	\begin{center}
	\includegraphics[scale = 0.5]{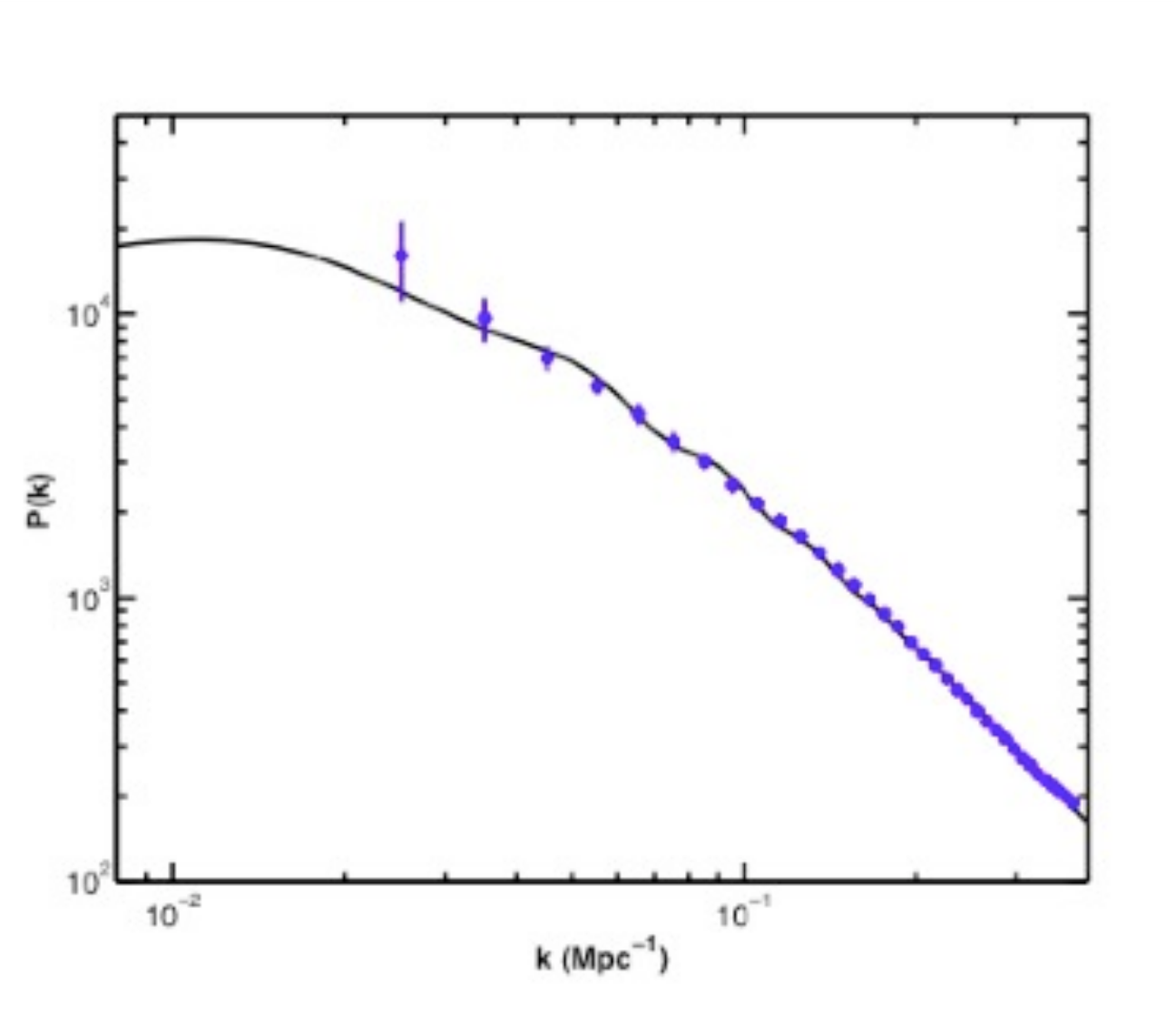}
	\includegraphics[scale = 0.4]{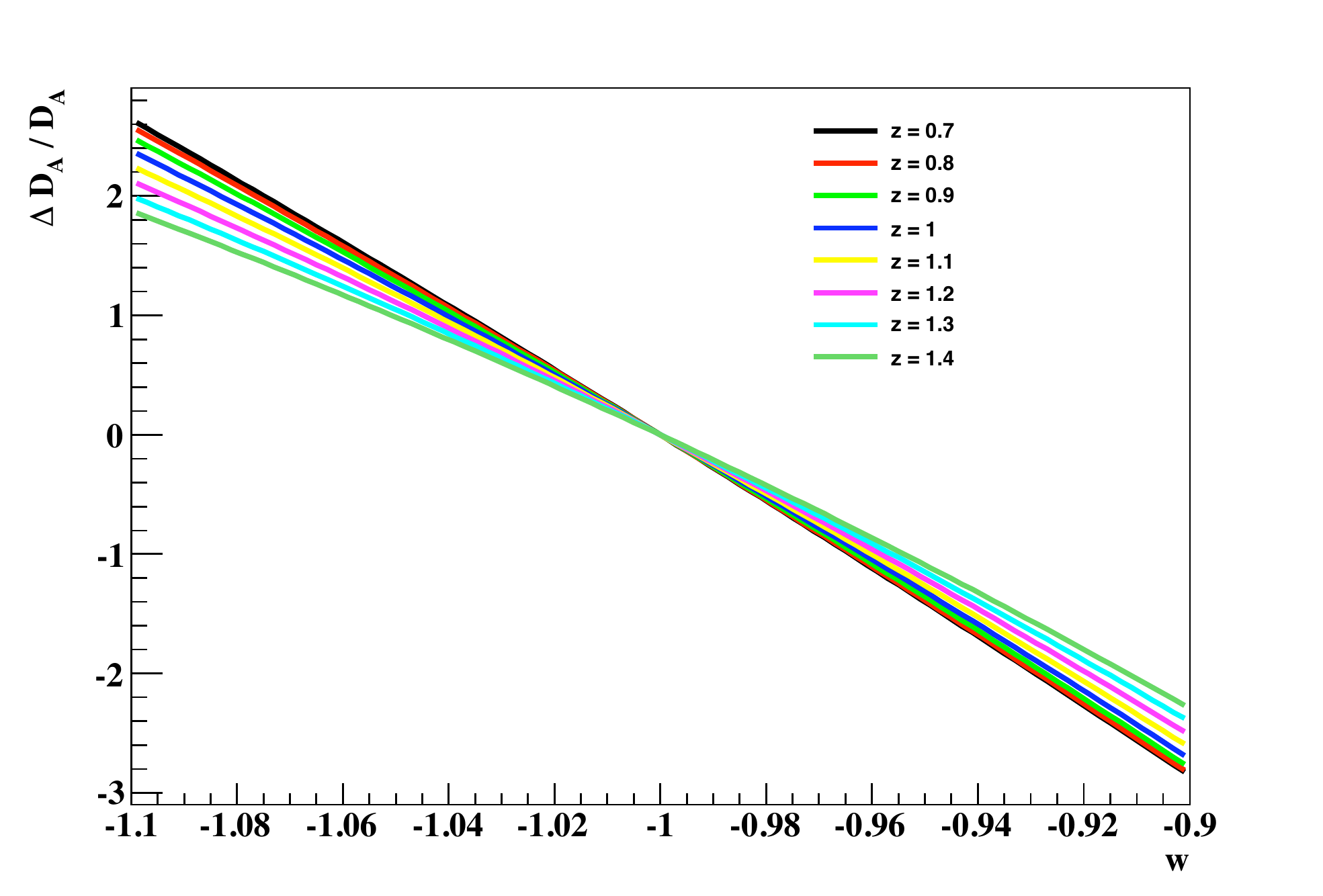}
	\caption{Left-hand panel: Reconstructed power spectrum for $z = 0.7$. Right-hand panel: Evolution of the relative uncertainties of the angular diameter distance $D_A$ as a function of $w$ and the redshift slice.}
	\label{Fig:Daw}
	\end{center}
\end{figure}

For a very preliminarysimulation of $\unit[5]{Gpc^3}$ and $z\in[0.7;1.4]$, we obtain a galaxy catalog of $2.6\times10^7$ galaxies. The BAO wavelength uncertainties computed by fitting the power spectrum shown Fig~\ref{Fig:Daw} on the right-hand panel is equal to $3\%$ at $z = 0,7$. Looking at the evolution of the relative uncertainty on $D_A$ as a function of $w$ and the redshift, gives a relative uncertainty on the dark energy parameter $\Delta w/w  = 13\%$ for $\Delta D_A/D_A = 3\%$.

\bibliography{biblio}
\bibliographystyle{unsrt}

\end{document}